\documentclass[aps,prl,twocolumn,superscriptaddress]{revtex4-1}
\usepackage{xspace}
\usepackage{bm,physics}
\usepackage{graphicx}
\usepackage{xcolor}
\usepackage{amsmath,amssymb,amsfonts,amsthm}
\usepackage[colorlinks=true,linkcolor=blue,anchorcolor=red,citecolor=blue,urlcolor=blue]{hyperref}
\usepackage{multirow}
\usepackage{diagbox}

\def \k {\bm{k}}

\def \H {\mathcal{H}}
\def \k {\bm{k}}

\begin{document}

\title{$PT$ Symmetry's Real Topology}

\author{J. X. Dai}
\affiliation{Department of Physics and HK Institute of Quantum Science \& Technology, The University of Hong Kong, Pokfulam Road, Hong Kong, China}

\author{Y. X. Zhao}
\email[]{yuxinphy@hku.hk}
\affiliation{Department of Physics and HK Institute of Quantum Science \& Technology, The University of Hong Kong, Pokfulam Road, Hong Kong, China}

\begin{abstract}
	Symmetry-protected topological phases have been a central theme in condensed matter physics and beyond over the past two decades. Most efforts have focused on topological classifications of physical systems under given symmetries, while the intrinsic topology of the symmetries themselves has received much less attention. Here, we show that, in generic non-interacting spinless crystals, the  spacetime inversion symmetry $PT$ naturally carries a real vector-bundle structure whose topology is characterized by Stiefel--Whitney (SW) classes. In contrast to previous work, where SW classes were used to describe the topology of real valence bundles protected by $PT$, we identify SW classes associated to the $PT$ symmetry itself. These symmetry SW classes can endow the \emph{total} real bundle of a $PT$-symmetric band structure with nontrivial topology, overturning the common assumption that the total bundle is always trivial. As a consequence, valence and conduction bands can exhibit asymmetric SW classes, in sharp contrast to the usual symmetric scenario. We further demonstrate that the symmetry SW classes provide a refined distinction between atomic insulator phases. Our results underscore the importance of treating crystal symmetries as topological objects in their own right, rather than focusing solely on the topology of energy bands.
\end{abstract}

\maketitle

{\color{blue}\textit{Introduction}} The past two decades have witnessed major progress in the study of symmetry-protected topological (SPT) phases in condensed matter and related fields, under the paradigm that configurations of matter are organized into distinct components by symmetry constraints~\cite{Hasan2010RMP,XLQi2011RMP,Schnyder2016RMP,Vishwanath2018RMP}. Among various symmetries, spinless $PT$ symmetry, the combination of spatial inversion $P$ and time reversal $T$, has recently attracted broad interest in electronic materials with negligible spin--orbit coupling and in artificial crystals~\cite{BJY2019PRX,ShengXL2019PRL,Tomas2019Science,YangEr2020PRL,Wang2020PRL,lee2020npj,guo2021nature,jiang2021NC,ShengXL2022PRL,Chen2021PRB,Zhu2022PRB,xue2023stiefel}. This is because it leads to real band structures characterized by Euler and Stiefel--Whitney (SW) classes~\cite{YXZHAO2016PRL,YXZHAO2017PRL,BJY2018PRL}, enabling higher-order topological insulators and semimetals realizable in both quantum materials and photonic or acoustic systems. 

Most previous works either a priori treated the $PT$ symmetry operator $\mathcal{PT}$ as independent of $\bm{k}$, or did not carefully address the $\bm{k}$-dependence of
\begin{equation}
	\mathcal{PT} = U_{PT}(\bm{k}) \mathcal{K},
\end{equation}
with $\mathcal{K}$ the complex conjugation~\cite{PT_notes}. However, when the inversion center is displaced from the unit-cell center---a ubiquitous situation---$\mathcal{PT}$ acquires an intrinsic $\bm{k}$-dependence whose topology, as we show, has important consequences for band topology.

To capture this topology, we consider the real vector bundle naturally induced by $\mathcal{PT}$. Generalizing the notion of symmetry beyond group theory is an active line of research in quantum many-body physics. Here we demonstrate that such generalization is also necessary in band theory for single-particle states: the symmetry framework must involve not only groups and their representations, but also fiber bundles.

The resulting real vector bundle is an intrinsic structure of $\mathcal{PT}$, originating from the Takagi decomposition of $U_{PT}(\bm{k})$~\cite{brezinski2022journey}. It is therefore fundamentally different from the widely studied real vector bundle of energy bands protected by $PT$ symmetry. While previous work focused on the topology of energy bands, we turn our attention to the topology of the symmetry operator itself.

The topology of the real vector bundle defined by $PT$ over the Brillouin zone is characterized by the first and second SW classes $w_i^{PT}$ with $i = 1,2$. Conventionally, for a given model, the total set of bands (valence plus conduction) is assumed to form a trivial bundle. By contrast, when $w_i^{PT}$ are nontrivial, $PT$ symmetry endows the entire real band bundle with the same nontrivial $w_i^{PT}$. As a consequence, the valence and conduction bundles can carry asymmetric SW classes, in sharp contrast to the standard symmetric scenario. Unlike ordinary band topological invariants, SW classes protected by a topologically nontrivial $PT$ symmetry do not necessarily give rise to in-gap boundary states. Moreover, the SW classes of $PT$ provide a refined distinction between atomic insulator phases. Although each atomic insulator is usually regarded as topologically trivial in a physical sense, phase transitions between them necessarily pass through topological semimetals due to the mismatch of their topological invariants, and the corresponding zero modes carry nontrivial SW topological charges.

We illustrate these features of topologically nontrivial $PT$ symmetry using explicit models: a one-dimensional lattice model, a two-dimensional square-lattice semimetal, a two-dimensional kagome-lattice insulator, and a three-dimensional cubic-lattice semimetal. These models may be relevant to quantum materials and are also realizable in artificial crystal platforms \cite{LuNaturePhotonics2014,YangPRL2015,HuberNaturePhysics2016,Ronny_2018np,Serra_Garcia_2018nature,Peterson_2018nature,Zhang2018Adip,OzawaRMP2019,MaGC_2019nature,Yu_Zhao_NSR2020,NiNatureCommu2020}.

We expect that our results will stimulate further investigations of how the topology of symmetry operators constrains and enriches the topology of quantum matter.

{\color{blue}\textit{$PT$ symmetry's real vector bundle}} 
We begin by examining the $\bm{k}$-dependence of $\mathcal{PT}$. 
When inversion $P$ does not preserve the unit cells, the inversion operator $\mathcal{P}$ necessarily acquires $\bm{k}$-dependence that cannot be removed by any redefinition of the unit cell. 
In a tight-binding model, this situation occurs when sites in a unit cell are mapped, under inversion, to sites belonging to more than one unit cell for all possible choices of unit cell. 
For example, if a primitive unit cell contains an even number of sites and the inversion center is located on a site, then inversion cannot preserve any choice of unit cell, because any set of sites invariant under inversion must contain an odd number of sites. 
In general, we therefore write
$\mathcal{PT} = U_{PT}(\bm{k})\mathcal{K}$, and there always exists a basis for each unit cell such that $U_{PT}(\bm{k})$ is periodic in the Brillouin zone, i.e., $U_{PT}(\bm{k})$ is a continuous map from the Brillouin torus $T^d_F$ to $\mathrm{U}(N)$.

For spinless systems, $(\mathcal{PT})^2 = 1$, or equivalently $U_{PT}(\bm{k})[U_{PT}(\bm{k})]^* = 1$. 
Since $U_{PT}(\bm{k})$ is unitary, we have $[U_{PT}(\bm{k})]^\dagger = [U_{PT}(\bm{k})]^*$. 
Thus, for each $\bm{k}$, $U_{PT}(\bm{k})$ is not only unitary but also symmetric, i.e.,
\begin{equation}
	U_{PT}(\bm{k}) [U_{PT}(\bm{k})]^\dagger = 1,\quad 
	U_{PT}(\bm{k}) = [U_{PT}(\bm{k})]^T,
\end{equation}
and will be referred to as a symmetric unitary matrix.

Any symmetric unitary matrix $M$ can be written as $M = U U^T$ with $U$ a unitary matrix, known as the Takagi factorization \cite{brezinski2022journey,DAI2021PRB,shiozaki2025mathbb}. 
The Takagi factor $U$ is not unique: $\tilde{U}$ is also a Takagi factor if and only if $\tilde{U}=UO$ for a unique real orthogonal matrix $O$, and clearly $M=\tilde{U}\tilde{U}^T$ since $OO^T=1$.

In general, it is not possible to choose a global Takagi factorization for $U_{PT}(\bm{k})$ over the entire Brillouin torus $T_F^d$. 
However, we can choose a finite open cover $T^d_F = \bigcup_{\alpha} X^{\alpha}$ such that on each patch $X^\alpha$ there exists a continuous Takagi factor $\mathcal{U}^\alpha(\bm{k})$ satisfying
\begin{equation}\label{eq:Takagifactor}
	U_{PT}(\bm{k}) = \mathcal{U}^\alpha(\bm{k})[\mathcal{U}^\alpha(\bm{k})]^T,
\end{equation}
for all $\bm{k}\in X^\alpha$. 
For two patches $X^\alpha$ and $X^\beta$ with nonempty intersection $X^{\alpha\beta} = X^\alpha \cap X^\beta$, their Takagi factors over $X^{\alpha\beta}$ are related by
\begin{equation}\label{eq:trans_U}
	\mathcal{U}^\alpha(\bm{k}) = \mathcal{U}^\beta(\bm{k})\, \mathcal{O}^{\beta\alpha}(\bm{k}),
\end{equation}
where $\mathcal{O}^{\beta\alpha}(\bm{k})$ is a continuous map from $X^{\alpha\beta}$ to the group of real orthogonal matrices $\mathrm{O}(N)$.

Furthermore, for any three patches $X^{\alpha}$, $X^{\beta}$, and $X^{\gamma}$ with nonempty triple overlap $X^{\alpha\beta\gamma} = X^\alpha \cap X^\beta \cap X^\gamma$, we have
\begin{equation}
	\mathcal{O}^{\alpha\gamma}(\bm{k})\, \mathcal{O}^{\gamma\beta}(\bm{k})\, \mathcal{O}^{\beta\alpha}(\bm{k}) = 1,
\end{equation}
for all $\bm{k}\in X^{\alpha\beta\gamma}$.

Therefore, the $\mathcal{O}^{\alpha\beta}$ can be interpreted as compatible transition functions. 
The cover $\{X^\alpha\}$ together with the transition functions $\{\mathcal{O}^{\alpha\beta}\}$ defined on the overlaps $X^{\alpha\beta}$ provide precisely the data specifying a real vector bundle (equivalently, a principal $\mathrm{O}(N)$ bundle) over the Brillouin torus $T_F^d$, and conversely any such bundle arises this way. 
We denote this real vector bundle by $E_{PT}$ and refer to it as the $PT$ symmetry's real vector bundle. 
The topology of $E_{PT}$ encodes the topology of $U_{PT}(\bm{k})$. 
For $d \le 3$, real vector bundles over $T^d_F$ are classified by their Stiefel-Whitney (SW) classes. 
If two $PT$ symmetry real vector bundles have different SW classes, then the corresponding $PT$ symmetry operators belong to distinct topological classes.

{\color{blue}\textit{$PT$ symmetry's real energy bands}} 
We now reveal how the SW classes of $PT$’s real vector bundle constrain the real bands of a $PT$-symmetric Hamiltonian $\mathcal{H}(\bm{k})$.

The complex band structure defines the trivial complex vector bundle $\mathbb{C}^N \times T_F^d$, which is equivalent to the trivial real vector bundle $\mathbb{R}^{2N} \times T_F^d$. 
The $PT$-symmetry operator $\mathcal{PT} = U_{PT}(\bm{k})\mathcal{K}$ selects, at each $\bm{k}$, an $N$-dimensional real subspace
$V_{PT}(\bm{k})$ consisting of vectors $\psi(\bm{k})$ that satisfy $U_{PT}(\bm{k})\psi(\bm{k})^* = \psi(\bm{k})$. 
Each $V_{PT}(\bm{k})$ is a real vector space, as it is closed under linear combinations with real coefficients. 
These $N$-dimensional subspaces vary continuously over $T_F^d$ and thus form an $N$-dimensional real vector bundle $V_{PT}$. 
As we now show, $V_{PT}$ is isomorphic to the $PT$ symmetry real vector bundle $E_{PT}$.

We continue to use the cover $T^d_F = \bigcup_{\alpha} X^{\alpha}$. 
On each patch $X^\alpha$, the operator $\mathcal{PT}$ can be transformed to pure complex conjugation by the Takagi factor:
\begin{equation}
	\widetilde{\mathcal{PT}}
	= [\mathcal{U}^\alpha(\bm{k})]^\dagger \, \mathcal{PT} \, \mathcal{U}^\alpha(\bm{k})
	= \mathcal{K}.
\end{equation}
For any $\psi(\bm{k}) \in V_{PT}(\bm{k})$, the transformed vector 
$\psi^\alpha(\bm{k}) = [\mathcal{U}^\alpha(\bm{k})]^\dagger \psi(\bm{k})$ is real, i.e., $\psi^\alpha(\bm{k}) \in \mathbb{R}^N$, because $\widetilde{\mathcal{PT}} \psi^\alpha = (\psi^\alpha)^* = \psi^\alpha$. 
Thus, $[\mathcal{U}^\alpha(\bm{k})]^\dagger$ provides a trivialization of $V_{PT}$ over $X^{\alpha}$, identifying it with $\mathbb{R}^N \times X^\alpha$. 
On the overlap $X^{\alpha\beta}$, the transition function between the trivializations on $X^\beta$ and $X^\alpha$ is exactly $\mathcal{O}^{\alpha\beta}(\bm{k})$, since
$\psi^\alpha(\bm{k}) = \mathcal{O}^{\alpha\beta}(\bm{k}) \psi^\beta(\bm{k})$ for all $\bm{k}\in X^{\alpha\beta}$ [Eq.~\eqref{eq:trans_U}]. 
Hence $V_{PT}$ is isomorphic to the $PT$ symmetry bundle $E_{PT}$. 
From now on, we do not distinguish between them and simply refer to both as the $PT$ symmetry’s real vector bundle $E_{PT}$.

\begin{figure*}
	\centering
	\includegraphics[width=\textwidth]{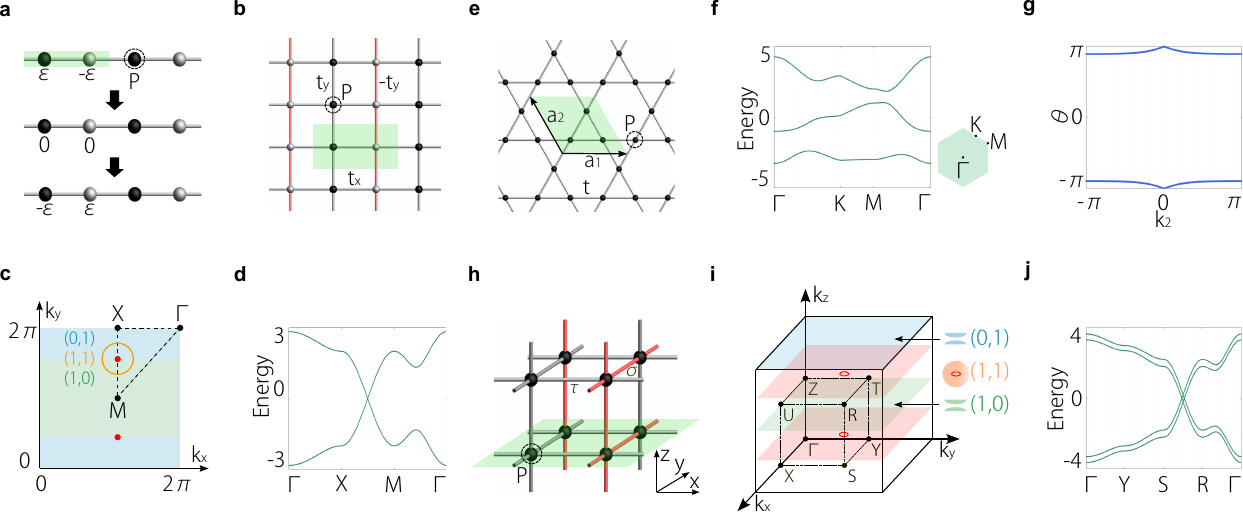}
	\caption{\textbf{a} Schematic of the 1D lattice model with nearest-neighbor hopping amplitude $t$. Sites of different colors represent sublattices with opposite onsite energies $\pm \epsilon$. This setup illustrates the evolution between distinct atomic insulator phases separated by a critical point. \textbf{b} Structure of the 2D semimetal on a square lattice. Gray (red) bonds denote positive (negative) hopping amplitudes. \textbf{c} The Brillouin zone featuring two nodal points (red dots). Each point carries a topological charge $(w_{1}^{+},w_{1}^{-})=(1,1)$ defined on an enclosing circle. \textbf{d} Bulk band structure of $\H(\mathbf{k})$ in Eq.~\eqref{eq:Hsquare} along high-symmetry lines. \textbf{e} The Kagome lattice with primitive vectors $\mathbf{a}_1, \mathbf{a}_2$. All nearest-neighbor hoppings are $t$, and $\epsilon_i$ denotes the onsite energy of the $i$-th site. \textbf{f} Bulk band structure for the model in \textbf{e} along the path $\Gamma-K-M-\Gamma$. \textbf{g} Wilson loop spectra for the conduction bands of Eq.~\eqref{eq:kagomeH}, calculated along circles $C_{k_2}$ ($k_2$ fixed) and parameterized by $k_1$, confirming $w_2^{+}=1$. \textbf{h} The 3D cubic lattice model with four sites per unit cell. Bond colors follow the convention in \textbf{b}. \textbf{i} Brillouin zone exhibiting two nodal loops (red circles). Each loop carries a charge $(w_2^{+},w_2^{-})=(1,1)$ defined on an enclosing sphere. \textbf{j} Bulk band structure for the model in \textbf{h} along the path $\Gamma-Y-S-R-\Gamma$.} \label{FIG1}
\end{figure*}

To connect with real-band theory, we note that on each patch $X^\alpha$ the Hamiltonian $\mathcal{H}(\bm{k})$ can be transformed into a real symmetric matrix by
\begin{equation}
	\mathcal{H}^\alpha(\bm{k})
	= [\mathcal{U}^\alpha(\bm{k})]^\dagger \mathcal{H}(\bm{k}) \mathcal{U}^\alpha(\bm{k})
	\in \mathrm{SM}_{\mathbb{R}}(N),
\end{equation}
for all $\bm{k}\in X^\alpha$, where $\mathrm{SM}_{\mathbb{R}}(N)$ denotes the space of real symmetric $N\times N$ matrices. 
The matrix $\mathcal{H}^\alpha(\bm{k})$ is real because $[\widetilde{\mathcal{PT}}, \mathcal{H}^\alpha(\bm{k})]=0$, and a real Hermitian matrix is simply a real symmetric matrix. 
On each overlap $X^{\alpha\beta}$, the real Hamiltonians $\mathcal{H}^\alpha(\bm{k})$ and $\mathcal{H}^\beta(\bm{k})$ are related by
\begin{equation}
	\mathcal{H}^\alpha(\bm{k})
	= [\mathcal{O}^{\beta\alpha}(\bm{k})]^T \mathcal{H}^\beta(\bm{k}) \, \mathcal{O}^{\beta\alpha}(\bm{k}),
\end{equation}
as follows from Eq.~\eqref{eq:trans_U}. 
Thus, real Bloch functions are precisely sections of $E_{PT}$.

If $E_{PT}$ is trivial (for instance, in the case $\mathcal{PT} = \mathcal{K}$), then for a $PT$-symmetric insulator the valence and conduction real vector bundles have identical SW classes, namely $w_i^+ = w_i^-$ for $i=1,2$. 
In general, however, $E_{PT}$ can have nontrivial SW classes $w_i^{PT}$, and hence
\begin{equation}
	w_i^+ + w_i^- = w_i^{PT},
\end{equation}
for $i=1,2$. 
Consequently, the valence and conduction bands can carry asymmetric SW classes.

{\color{blue}\textit{Lattice models}}
We now proceed to demonstrate our theory by concrete models with noncentered $PT$ symmetry. We will represent SW classes by SW numbers as introduced in the note~\cite{SW_notes}.

We start with the simplest $1$D model with equal nearest neighbor hopping amplitudes but different on-site energies $\pm \epsilon$ for two sublattices as illustrated in Fig.~\ref{FIG1}\textbf a. Each unit cell consists of two sites, and the inversion center is chosen as in Fig.~\ref{FIG1}\textbf a. Therefore, the unitary operator for  $PT$ symmetry is represented as
\begin{equation}\label{eq:1D_PT}
	U_{PT}(\bm{k})= \begin{bmatrix} 1 & 0 \\ 0 & e^{\text{i}k_x} \end{bmatrix}
\end{equation}
with $U_{PT}(\bm{k})U_{PT}(\bm{k})^*=1_2$. Then, the Takagi factor can be chosen as $\mathcal{U}(k_x)=\mathrm{diag}(1, e^{\text{i}k_x/2})$. The transition function attaching $\mathcal{U}(\pm\pi)$ is just $\mathcal{O}=\mathrm{diag}(1, -1)$ with $\det \mathcal{O}=-1$. Thus, $PT$-symmetry's bundle has nontrivial $w_1^{PT}=1$, which means that for the conduction and valence bands $(w_1^{+}, w_1^{-})=(1,0)$ or $(0,1)$. For the model, $w_1^{-}$ has been calculated previously. 

Whether $w_1^{\pm}=0$ or $1$ depends on whether the conduction/valence band is dominated by the $A$-sublattice or $B$-sublattice. Both cases correspond to atomic insulators and therefore have no end states \cite{bradlyn2017topological,po2017symmetry,schnyder2018quantized}. But, since they have distinct $w_1^{\pm}$, they belong to distinct atomic phases and there exists a critical point between them.

The critical point is manifested as a Dirac semimetal, analogous to graphene with two Dirac points, if the $1$D systems with distinct SW classes are realized as $k_x$-subsystems for a $2$D system. Such a model is given by
\begin{equation}\label{eq:Hsquare}
	\mathcal{H}(\bm{k}) = t_x(1+\cos k_x)\sigma_1 + t_x\sin k_x\sigma_2 + 2t_y\cos k_y\sigma_3,
\end{equation}
as illustrated in Fig.~\ref{FIG1}\textbf b, and $PT$ symmetry is still represented by \eqref{eq:1D_PT}. With $t_x=t_y=1$, the model hosts two twofold degenerate Dirac points at $\bm{K}_{\pm}=(\pi,\pm\pi/2)$, which divide the sub $k_x$-insulators into two regions with $k_y\in (-\pi/2,\pi/2)$ or $(-\pi, -\pi/2)\cup (\pi/2,\pi]$, as marked in green and blue, respectively in Fig.~\ref{FIG1}\textbf c.  $(w_{1,x}^{+},w_{1,x}^{-})=(1,0)$ and $(0,1)$ in the two regions, respectively. As a result, $(w_{1,x}^{+},w_{1,x}^{-})$ on a local circle surrounding a Dirac point are $(1,1)$, equal to the difference of $(w_{1,x}^{+},w_{1,x}^{-})$ in the two regions. Thus, locally around each Dirac point, the topological band structure is exactly the same as that for a Dirac point in graphene. However, as each sub $k_x$-insulator is essentially atomic, there exist no topologically protected in-gap edge states in contrast to the case of graphene.

\textit{Kagome insulator} Next, we consider a kagome lattice model, 
\begin{equation}\label{eq:kagomeH}
	\mathcal{H}(\bm{k}) = \begin{bmatrix}
		\epsilon_1 & q_1(\bm{k}) & q_2(\bm{k}) \\
		q_1^*(\bm{k}) & \epsilon_2 & q_3(\bm{k}) \\
		q_2^*(\bm{k}) & q_3^*(\bm{k}) & \epsilon_3
	\end{bmatrix},
\end{equation}
with three sites per unit cell, as shown in Fig.~\ref{FIG1}\textbf e. 
Here, $q_1(\bm{k}) = t(1+e^{-\text{i}k_1})$, $q_2(\bm{k}) = t(1+e^{-\text{i}(k_1+k_2)})$, and $q_3(\bm{k}) = t(1+e^{-\text{i}k_2})$.

All hopping amplitudes are real, preserving time-reversal symmetry, and the inversion center is located at a lattice site away from the unit cell center, leading to 
\begin{equation}
	U_{PT}(\bm{k}) = \begin{bmatrix} 1 & 0 & 0\\ 0 & e^{\text{i}k_1} & 0\\ 0 & 0 & e^{\text{i}(k_1+k_2)} \end{bmatrix},
\end{equation}
where $k_i=\bm{k}\cdot \bm{a}_i$ with $\bm{a}_i$ two primitive lattice vectors. The Takagi factorization can be chosen as $\mathcal{U}(\bm{k})=\mathrm{diag}(1, M(\bm{k}))$, with
\begin{equation}\label{eq:M}
M(k_1,k_2)=\begin{bmatrix}
		\cos \frac{k_1}{2} e^{\text{i}\frac{k_1}{2}} & \sin \frac{k_1}{2} e^{\text{i}\frac{k_1}{2}}\\
		-\sin\frac{k_1}{2} e^{\text{i}\frac{k_1+k_2}{2}} & \cos \frac{k_1}{2} e^{\text{i}\frac{k_1+k_2}{2}}
	\end{bmatrix}.
\end{equation}
It is straightforward to check that $\mathcal{U}(k_x+2\pi,k_y)=\mathcal{U}(k_x,k_y)$. Hence, we introduce the transition function,
\begin{equation}
	\mathcal{O}(k_1) = \begin{bmatrix}
		1 & 0 & 0\\0 &\cos k_1 & \sin k_1 \\ 0 & \sin k_1 & -\cos k_1
	\end{bmatrix},
\end{equation}
so that $\mathcal{U}(k_x,0)=\mathcal{U}(k_x,2\pi)\mathcal{O}(k_x)$. The transition function implies that $w^{PT}_{1, \bm{a}_1}=0$, $w^{PT}_{1,\bm{a}_2}=1$, and $w^{PT}_2=1$, since $\det \mathcal{O}=-1$ and $\mathcal{O}(k_1)$ has a unit winding number over $\mathrm O(2)\subset \mathrm O(3)$. See the SM for more details~\cite{Supp}.

With $\epsilon_1 = -2$, $\epsilon_2 = \epsilon_3 = t = 1$, the system has two conduction bands and one valence band, as illustrated in Fig.~\ref{FIG1}\textbf f. While the single valence band clearly carries trivial $w^{-}_2$, the Wilson-loop calculation in Fig.~\ref{FIG1}\textbf g shows $w_2^{+}=1$ for the two conduction bands~\cite{Yue2024PRB,Supp}. This asymmetric distribution $(w_2^{+}, w_2^{-})=(1,0)$ is consistent with $w^{PT}_2=1$. In contrast to the conventional SW insulator, this insulator is atomic without edge and corner in-gap states. This can be seen through the atomic limit $t\rightarrow 0$, which preserves the energy gap and therefore the SW numbers.

\textit{3D SW semimetal} The last example is a 3D cubic lattice model with $\pi$-flux through each plaquette along any direction, as illustrated in Fig.~\ref{FIG1}\textbf h.  Each unit cell contains $2\times 2$ sites, on which two sets of the standard Pauli matrices $\sigma$ and $\tau$ act, and all hopping amplitudes are real, preserving time-reversal symmetry. Their magnitudes are denoted by $t_i$ with $i=x,y,z$, and signs $\pm$ are indicated by gray and red, respectively. The Hamiltonian is in the Dirac form,
\begin{equation}\label{eq:Hamilcube}
	\mathcal{H}_0(\k)=\sum_{a=1}^5 f_a(\bm{k})\gamma^a.
\end{equation}
Here, $f_1(\bm{k})=t_x(1+\cos k_x)$, $f_2(\bm{k})=t_x\sin k_x$, $f_3(\bm{k})=t_y(1+\cos k_y)$, $f_4(\bm{k})=t_y\sin k_y$, $f_5(\bm{k})=2t_z\cos k_z$, and the Dirac matrices are given by
$\gamma^{1}=\tau_1\otimes\sigma_0$, $\gamma^{2}=\tau_2\otimes\sigma_0$, $\gamma^{3}=\tau_3\otimes\sigma_1$, $\gamma^{4}=\tau_3\otimes\sigma_2$,  $\gamma^{5}=\tau_3\otimes\sigma_3$, satisfying $\{\gamma^a,\gamma^b\}=2\delta_{ab}1_4$.

The inversion center at a site leads to 
\begin{equation}
	U_{PT}(\bm{k}) = \begin{bmatrix} 1 & 0 & 0 & 0 \\ 0 & e^{\text{i}k_y} & 0 & 0 \\ 0 & 0 & e^{\text{i}k_x} & 0 \\ 0 & 0 & 0 & e^{\text{i}(k_x+k_y)} \end{bmatrix}.
\end{equation}
Observing the similarity with the previous model, we choose the Takagi factorization in the block diagonal form,
$
	\mathcal{U}(\bm{k}) = \mathrm{diag}(1, e^{\text{i}\frac{k_y}{2}}, M(k_x,k_y))
$
with $M(k_x,k_y)$ given by \eqref{eq:M}. Since $\mathcal{U}(k_x+2\pi,k_y)=\mathcal{U}(k_x,k_y)$, we need to introduce the transition function,
\begin{equation}
	\mathcal{O}(k_x) = \begin{bmatrix}
		1 & 0 & 0 & 0\\0 &-1 & 0 & 0 \\ 0 & 0 & \cos k_x & \sin k_x \\ 0 & 0 & \sin k_x & -\cos k_x
	\end{bmatrix},
\end{equation}
to attach $\mathcal{U}(k_x,0)$ and $\mathcal{U}(k_x,2\pi)$, i.e., $\mathcal{U}(k_x,0)=\mathcal{U}(k_x,2\pi)\mathcal{O}(k_x)$. Since $\det \mathcal{O}(k_x)=1$, $w_1^{PT}$ is trivial. But, from the unit winding number of $\mathcal{O}(k_x)$ in the subgroup $\mathrm O(2)\subset\mathrm O(4)$,  we see $w^{PT}_{2,xy}=1$, implying asymmetric $w^{\pm}_{2,xy}$ for $k_x$-$k_y$ insulators.

The Hamiltonian~\eqref{eq:Hamilcube} describes a 3D real Dirac semimetal with two fourfold real Dirac points at $\bm
K_{\pm}=(\pi,\pi,\pm\pi/2)$ for $t_i>0$. We can add a $PT$-invariant perturbation term $\Delta=\lambda\,\tau_0\otimes\sigma_3$, i.e.,
$\mathcal{H} = \mathcal{H}_0+ \Delta$. Then, each real Dirac point is spread into a nodal loop as illustrated in Fig.~\ref{FIG1}\textbf i~\cite{Wang2020PRL,Shao2021PRL}. 

On any local sphere surrounding each nodal loop, the second SW charges are given by $(w_2^+,w_2^{-})=(1,1)$. Thus, locally in momentum space, the nodal loop here is exactly the same as the previously studied four-band nodal loop spread from a real Dirac point. 

However, the non-centered $PT$-symmetric nodal-loop semimetal has no in-gap hinge states that are induced from the SW charges in the centered $PT$-symmetry case, as indicated in Fig.~\ref{FIG1}\textbf h. This is because the sub $k_x$-$k_y$ insulators on the two sides of each nodal loop or Dirac point are essentially atomic insulators (similar to the previous kagome case), although they have distinct asymmetric SW numbers $(w_2^{+},w_2^{-})=(1, 0)$ and $(0,1)$ as shown in Fig.~\ref{FIG1}\textbf i.

{\color{blue}\textit{Summary and discussions}} The conventional paradigm of SPT phases is that the configurations of a physical system are topologically classified under the constraints of a given symmetry group. Our work on the $PT$ symmetry's real topology suggests an extension of this paradigm for band theory: first, one topologically classifies the symmetry group; then, for a given topology of the symmetry group, one follows the conventional paradigm to further topologically classify the configurations of the system. The topology of the symmetry group in momentum space fundamentally affects the topological classification of energy bands, just as the monopole charge of the background gauge group is crucial for the behavior of fermions in Yang–Mills theory~\cite{t1974magnetic,polyakov1974particle}.

Another remarkable feature of this work is about the topological distinction among atomic insulators, which are typically considered trivial in conventional topological band theory \cite{bradlyn2017topological,po2017symmetry}. Constrained by topologically nontrivial $PT$ symmetry, different atomic insulators are distinguished by their asymmetric Stiefel-Whitney classes and cannot be adiabatically connected. As a result, their transition points are topologically nontrivial and can lead to topological semimetals with conventional SW topological charges. But in contrast to conventional topological semimetals, these semimetals do not have in-gap boundary states as the sub-insulators are all atomic.

Finally, it is worth mentioning that there exist mixed cases, i.e.,  $PT$ symmetry's real bundle is topologically nontrivial along some dimensions, while trivial along the other dimensions. Then, conventional $PT$-symmetric topological band theory still applies along these trivial dimensions of $PT$ symmetry.  For instance, in the graphite model in Ref.~\cite{Shao2021PRL}, only $w_{1,z}^{PT}$ is nontrivial, and therefore $w_{2,xy}$ is unaffected. Consequently, hinge states are observed in the  SW nodal-line semimetal.

\bibliographystyle{apsrev4-1}
\bibliography{references}

\begin{thebibliography}{43}
\bibitem [{PT_()}]{PT_notes}%
  \BibitemOpen
  \href@noop {} {}\bibinfo {note} {Here, the operator is the $\mathcal{PT}$
  operator for the whole system, not just for valence bands of an
  insulator.}\BibitemShut {Stop}%
\bibitem [{SW_()}]{SW_notes}%
  \BibitemOpen
  \href@noop {} {}\bibinfo {note} {For a $1$D system, $w_1$ is valued in
  $H^1(T_F^1, \mathbb{Z}_2)\cong \mathbb{Z}_2$ with $T^1_F$ the Brillouin
  1-torus. $w_1$ is completely determined by its value $w_1(T_F^1)$ on $T_F^1$,
  called the first SW number, and therefore $w_1$ is used to denote both the
  first SW class and the first SW number. For a $2$D system, $w_1$ is valued in
  $H^1(T^2_F,\mathbb{Z}_2)\cong \mathbb{Z}_2^2$ with $T_F^2$ the Brillouin
  2-torus. $w_1$ can be represented by the first SW numbers
  $w_1(\gamma_{i})=w_{1,i}$ on the two large circles $\gamma_{i}$ with $i=1,2$.
  $w_2\in H^2(T^2_F,\mathbb{Z}_2)\cong \mathbb{Z}_2$ is represented by the
  second SW number $w_2=w_2(T_F^2)$. For a $3$D system, $w_1\in
  H^1(T_F^3,\mathbb{Z}_2)\cong \mathbb{Z}_2^3$ is represented by the first SW
  numbers $w_{1,i}$ for its values on the three large circles $\gamma_i$ with
  $i=1,2,3$, and $ w_2\in H^2(T_F^3,\mathbb{Z}_2)\cong \mathbb{Z}_2^3$ by the
  second SW numbers $w_{2, ij}$ for its values on the three sub 2-tori labeled
  by $ij$.}\BibitemShut {Stop}%
\end{thebibliography}
\end{document}


\title{The Supplemental Materials for ``$PT$ Symmetry's Real Topology''}
	
	\author{J. X. Dai}
	\affiliation{Department of Physics and HKU-UCAS Joint Institute for Theoretical
		and Computational Physics at Hong Kong, The University of Hong Kong,
		Pokfulam Road, Hong Kong, China}

	\author{Y. X. Zhao}
	\email[]{yuxinphy@hku.hk}
	\affiliation{Department of Physics and HKU-UCAS Joint Institute for Theoretical
		and Computational Physics at Hong Kong, The University of Hong Kong,
		Pokfulam Road, Hong Kong, China}
	\maketitle
	%

	\setcounter{page}{1}
	\makeatletter
	\renewcommand\theequation{S\arabic{equation}}
	\renewcommand\thefigure{S\arabic{figure}}
	\setcounter{equation}{0}
	
	\section{The Stiefel-Whitney classes of the non-centered $PT$ symmetry operator}
	
	In crystalline systems, the inversion center often does not coincide with the unit cell center. This spatial offset leads to a $\k$-dependent $PT$ symmetry operator, $\mathcal{PT}=U_{PT}(\k)\mathcal{K}$. As detailed in the main text, $U_{PT}(\k)$ is a symmetric unitary matrix that admits the Takagi factorization
	\begin{equation}\label{eq:Takagi}
		U_{PT}(\k)=\U(\k)\U^T(\k), \quad \U\in\mathrm{U}(M+N).
	\end{equation}
	Here, $M$ and $N$ denote the number of valence and conduction bands, respectively. There is a gauge degree of freedom in this decomposition. Specifically, $U_{PT}$ remains invariant under the transformation:
	\begin{equation}
		\U(\k)\mapsto \U(\k)\O(\k), \quad \O\in\mathrm{O}(M+N).
	\end{equation}
	Consequently, the classifying space of the $\mathcal{PT}$ symmetry operator is the quotient space
	\begin{equation}
		US(M+N)=\mathrm{U}(M+N)/\mathrm{O}(M+N).
	\end{equation}
	The first and second Stiefel-Whitney classes originate from obstructions to a globally well-defined Takagi factorization $\U(\k)$, corresponding to the homotopy groups $\pi_0[\mathrm{O}(M+N)]\cong\mathbb{Z}_2$ and $\pi_1[\mathrm{O}(M+N)]\cong\mathbb{Z}_2$, respectively.
	
	To define the first Stiefel-Whitney class $w^{PT}_1$, we consider the base space as a circle $S^1$, divided into north and south hemicircles $D_{N,S}^1$ intersecting at $S^0=\{0,\pi\}$. We denote the Takagi factorization on these patches as $\U_{N}$ and $\U_{S}$. The transition function on the equator is then given by
	\begin{equation}\label{eq:transistion1}
		t(0/\pi)=\U_N^\dagger(0/\pi)\U_S(0/\pi)\in \mathrm{O}(M+N).
	\end{equation}
	The topological invariant $w_1^{PT}$ is obtained via the determinant relation: $(-1)^{w_1^{PT}}=\det[t(0)t(\pi)]$.
	
	For the second Stiefel-Whitney class $w^{PT}_2$, we consider the base space as a sphere $S^2$, divided into north and south hemispheres $D_{N,S}^2$ with the intersection $S^1$ parameterized by the angle $\phi$. The transition function on this equatorial $S^1$ is
	\begin{equation}\label{eq:transistion2}
		t(\phi)=\U_N^\dagger|_{S^1}(\phi)\U_S|_{S^1}(\phi)\in \mathrm{O}(M+N).
	\end{equation}
	The topological invariant $w_2^{PT}$ corresponds to the homotopy invariant of the mapping $t(\phi)$.
	
	\section{The Stiefel-Whitney classes of the bands}
	
	In this section, we detail the calculation of the Stiefel-Whitney classes for the valence and conduction bands under $\k$-dependent $PT$ symmetry.
	
	\subsection{Theoretical calculation methods}
	For a non-centered $PT$-symmetric system, we apply the unitary transformation $\U(\k)$ from Eq.~\eqref{eq:Takagi} to both the Hamiltonian $\H(\k)$ and the symmetry operator $\mathcal{PT}=U_{PT}(\k)\mathcal{K}$. This yields
	\begin{equation}\label{eq:transsys}
		\widetilde{\mathcal{P}}\widetilde{\mathcal{T}}=\U^\dagger(\k)\mathcal{PT}\U(\k)=\mathcal{K}, \quad \widetilde\H(\k)=\U^\dagger(\k)\H(\k)\U(\k).
	\end{equation}
	Under this transformation, $\mathcal{PT}$ is mapped to the complex conjugation operator $\mathcal{K}$, and $\H(\k)$ is mapped to a real Hamiltonian $\widetilde\H(\k)$. We can then decompose the transformed Hamiltonian as
	\begin{equation}
		\widetilde\H(\k)=\O(\k)\begin{bmatrix}
			1_M&\\&-1_N
		\end{bmatrix}\O^T(\k).
	\end{equation}
	Here, we have assumed without loss of generality that the eigenvalues for the conduction and valence bands are $+1$ and $-1$, respectively, which is sufficient for deriving the topology. Thus, the original Hamiltonian $\H(\k)$ can be expressed as
	\begin{equation}
		\H(\k)=\U(\k)\O(\k)\begin{bmatrix}
			1_M&\\&-1_N
		\end{bmatrix}\O^T(\k)\U^\dagger(\k).
	\end{equation}
	
	For the first Stiefel-Whitney class $w^{\pm}_1$, considering the base space $S^1$ divided into $D_{N,S}^1$, the transition functions $t_{\pm}(0/\pi)\in\mathrm{O}(M/N)$ for the conduction/valence bands on the equator are:
	\begin{equation}\label{eq:transistionpm1}
		\begin{bmatrix}
			t_+(0/\pi)&\\&t_-(0/\pi)
		\end{bmatrix}=\O_N^T(0/\pi)\U_N^\dagger(0/\pi)\U_S(0/\pi)\O_S(0/\pi).
	\end{equation}
	The invariant is determined by $(-1)^{w_1^{\pm}}=\det[t_\pm(0)t_\pm(\pi)]$.
	
	Similarly, for the second Stiefel-Whitney class $w^{\pm}_2$ on $S^2$, the transition functions $t_{\pm}(\phi)$ on the equator are given by:
	\begin{equation}\label{eq:transistionpm2}
		\begin{bmatrix}
			t_+(\phi)&\\&t_-(\phi)
		\end{bmatrix}=\O_N^T|_{S^1}(\phi)\U_N^\dagger|_{S^1}(\phi)\U_S|_{S^1}(\phi)\O_S|_{S^1}(\phi).
	\end{equation}
	The topological invariant $w_2^{\pm}$ is the homotopy invariant of $t_{\pm}(\phi)$.
	
	\subsection{Numerical calculation methods}
	We now present the numerical method for characterizing the band topology protected by $\k$-dependent $PT$ symmetry, as elaborated in Ref.~[39] of the main text. We introduce the numerical twisted Wilson loop operator for the valence bands, from which the Stiefel-Whitney classes are derived. The procedure for conduction bands is identical, differing only in the selection of bands.
	
	Consider a 2D insulator where the 1D sub-Brillouin zone (BZ), such as along $k_y$, is discretized into $n$ intervals. The discretization points are labeled by $i=0,1,\cdots,n-1$. The numerical twisted Wilson loop operator $W^{(\Lambda)}(k_x)$ is defined as 
	\begin{equation}
		W^{(\Lambda)}(k_x)=\lim_{n\to\infty}\left(\prod_{i=0}^{n-2} F_{i,i+1}\right)\times F_{n-1,n}^{(\Lambda)},
	\end{equation}
	where the overlap matrices are 
	\begin{equation}
		[F_{i,i+1}]_{ab}=\langle a,k_i|b,k_{i+1}\rangle, \quad [F_{n-1,n}^{(\Lambda)}]_{ab}=\langle a,k_{n-1}|\Lambda|b,k_0\rangle.
	\end{equation}
	Here, $W^{(\Lambda)}(k_{x})$ is calculated along a loop $C_{k_{x}}$ (with fixed $k_x$) parameterized by $k_y$. The state $|a,k_i\rangle$ represents the $a$-th valence band of the real Hamiltonian $\widetilde \H(k_x,k_i)$ [Eq.~\eqref{eq:transsys}]. $\Lambda$ serves as the transition function of the Takagi factorization $\U(\k)$ [Eq.~\eqref{eq:Takagi}] across the BZ boundary 
	\begin{equation}
		\Lambda=\mathcal{U}(k_x,k_y+2\pi) \mathcal{U}^\dagger(k_x,k_y).
	\end{equation}
	The real Hamiltonian satisfies the boundary condition $\widetilde\H(k_x,k_y+2\pi)=\Lambda\widetilde\H(k_x,k_y)\Lambda^\dagger$. 
	
	Let the eigenvalues of $W^{(\Lambda)}(k_x)$ be $e^{i\theta_a(k_x)}$ for $a=1,2,\cdots,N$. These eigenvalues are gauge invariant. The first Stiefel-Whitney class $w_{1,y}$ is obtained via 
	\begin{equation}
		(-1)^{w_{1,y}}=\det[W^{\Lambda}(k_x)]=\prod_{a=1}^N e^{i\theta_a(k_x)}.
	\end{equation}
	As for $w_{1,x}$, we can simply calculate the twisted Wilson loop operator $W^{(\Lambda)}(k_y)$ along the $k_x$-direction. Finally, $w_2$ is determined by the parity of the number of linear crossing points in the Wilson loop spectrum at $\theta=\pi$.